\documentstyle[12pt]{article}

\catcode`\@=11
\long\def\@makefntext#1{ 
\protect\noindent \hbox to 3.2pt {\hskip-.9pt
$^{{\ninerm\@thefnmark}}$\hfil}#1\hfill} 

\def\thefootnote{\fnsymbol{footnote}}
 \def\@makefnmark{\hbox to 0pt{$^{\@thefnmark}$\hss}}  

\def\ps@myheadings{\let\@mkboth\@gobbletwo
\def\@oddhead{\hbox{} 
\rightmark\hfil\ninerm\thepage}
\def\@oddfoot{}\def\@evenhead{\ninerm\thepage\hfil 
\leftmark\hbox{}}\def\@evenfoot{}
\def\sectionmark##1{}\def\subsectionmark##1{}}

\begin{document}

\newcommand{\symbolfootnote}{\renewcommand{\thefootnote}
	{\fnsymbol{footnote}}}
\renewcommand{\thefootnote}{\fnsymbol{footnote}}
\newcommand{\alphfootnote}
	{\setcounter{footnote}{0}
	 \renewcommand{\thefootnote}{\sevenrm\alph{footnote}}}

\newcounter{sectionc}\newcounter{subsectionc}\newcounter{subsubsectionc}
\renewcommand{\section}[1] {\vspace{0.6cm}\addtocounter{sectionc}{1}
\setcounter{subsectionc}{0}\setcounter{subsubsectionc}{0}\noindent
	{\bf\thesectionc. #1}\par\vspace{0.4cm}}
\renewcommand{\subsection}[1] {\vspace{0.6cm}\addtocounter{subsectionc}{1}
	\setcounter{subsubsectionc}{0}\noindent
	{\it\thesectionc.\thesubsectionc. #1}\par\vspace{0.4cm}}
\renewcommand{\subsubsection}[1]
{\vspace{0.6cm}\addtocounter{subsubsectionc}{1}
	\noindent {\rm\thesectionc.\thesubsectionc.\thesubsubsectionc.
	#1}\par\vspace{0.4cm}}
\newcommand{\nonumsection}[1] {\vspace{0.6cm}\noindent{\bf #1}
	\par\vspace{0.4cm}}

\newcounter{appendixc}
\newcounter{subappendixc}[appendixc]
\newcounter{subsubappendixc}[subappendixc]
\renewcommand{\thesubappendixc}{\Alph{appendixc}.\arabic{subappendixc}}
\renewcommand{\thesubsubappendixc}
	{\Alph{appendixc}.\arabic{subappendixc}.\arabic{subsubappendixc}}

\renewcommand{\appendix}[1] {\vspace{0.6cm}
        \refstepcounter{appendixc}
        \setcounter{figure}{0}
        \setcounter{table}{0}
        \setcounter{equation}{0}
        \renewcommand{\thefigure}{\Alph{appendixc}.\arabic{figure}}
        \renewcommand{\thetable}{\Alph{appendixc}.\arabic{table}}
        \renewcommand{\theappendixc}{\Alph{appendixc}}
        \renewcommand{\theequation}{\Alph{appendixc}.\arabic{equation}}
        \noindent{\bf Appendix \theappendixc #1}\par\vspace{0.4cm}}
\newcommand{\subappendix}[1] {\vspace{0.6cm}
        \refstepcounter{subappendixc}
        \noindent{\bf Appendix \thesubappendixc. #1}\par\vspace{0.4cm}}
\newcommand{\subsubappendix}[1] {\vspace{0.6cm}
        \refstepcounter{subsubappendixc}
        \noindent{\it Appendix \thesubsubappendixc. #1}
	\par\vspace{0.4cm}}

\def\abstracts#1{{
	\centering{\begin{minipage}{30pc}\tenrm\baselineskip=12pt\noindent
	\centerline{\tenrm ABSTRACT}\vspace{0.3cm}
	\parindent=0pt #1
	\end{minipage} }\par}}

\newcommand{\bibit}{\it}
\newcommand{\bibbf}{\bf}
\renewenvironment{thebibliography}[1]
	{\begin{list}{\arabic{enumi}.}
	{\usecounter{enumi}\setlength{\parsep}{0pt}
\setlength{\leftmargin 1.25cm}{\rightmargin 0pt}
	 \setlength{\itemsep}{0pt} \settowidth
	{\labelwidth}{#1.}\sloppy}}{\end{list}}

\topsep=0in\parsep=0in\itemsep=0in
\parindent=1.5pc

\newcounter{itemlistc}
\newcounter{romanlistc}
\newcounter{alphlistc}
\newcounter{arabiclistc}
\newenvironment{itemlist}
    	{\setcounter{itemlistc}{0}
	 \begin{list}{$\bullet$}
	{\usecounter{itemlistc}
	 \setlength{\parsep}{0pt}
	 \setlength{\itemsep}{0pt}}}{\end{list}}

\newenvironment{romanlist}
	{\setcounter{romanlistc}{0}
	 \begin{list}{$($\roman{romanlistc}$)$}
	{\usecounter{romanlistc}
	 \setlength{\parsep}{0pt}
	 \setlength{\itemsep}{0pt}}}{\end{list}}

\newenvironment{alphlist}
	{\setcounter{alphlistc}{0}
	 \begin{list}{$($\alph{alphlistc}$)$}
	{\usecounter{alphlistc}
	 \setlength{\parsep}{0pt}
	 \setlength{\itemsep}{0pt}}}{\end{list}}

\newenvironment{arabiclist}
	{\setcounter{arabiclistc}{0}
	 \begin{list}{\arabic{arabiclistc}}
	{\usecounter{arabiclistc}
	 \setlength{\parsep}{0pt}
	 \setlength{\itemsep}{0pt}}}{\end{list}}

\newcommand{\fcaption}[1]{
        \refstepcounter{figure}
        \setbox\@tempboxa = \hbox{\tenrm Fig.~\thefigure. #1}
        \ifdim \wd\@tempboxa > 6in
           {\begin{center}
        \parbox{6in}{\tenrm\baselineskip=12pt Fig.~\thefigure. #1 }
            \end{center}}
        \else
             {\begin{center}
             {\tenrm Fig.~\thefigure. #1}
              \end{center}}
        \fi}

\newcommand{\tcaption}[1]{
        \refstepcounter{table}
        \setbox\@tempboxa = \hbox{\tenrm Table~\thetable. #1}
        \ifdim \wd\@tempboxa > 6in
           {\begin{center}
        \parbox{6in}{\tenrm\baselineskip=12pt Table~\thetable. #1 }
            \end{center}}
        \else
             {\begin{center}
             {\tenrm Table~\thetable. #1}
              \end{center}}
        \fi}

\def\@citex[#1]#2{\if@filesw\immediate\write\@auxout
	{\string\citation{#2}}\fi
\def\@citea{}\@cite{\@for\@citeb:=#2\do
	{\@citea\def\@citea{,}\@ifundefined
	{b@\@citeb}{{\bf ?}\@warning
	{Citation `\@citeb' on page \thepage \space undefined}}
	{\csname b@\@citeb\endcsname}}}{#1}}

\newif\if@cghi
\def\cite{\@cghitrue\@ifnextchar [{\@tempswatrue
	\@citex}{\@tempswafalse\@citex[]}}
\def\citelow{\@cghifalse\@ifnextchar [{\@tempswatrue
	\@citex}{\@tempswafalse\@citex[]}}
\def\@cite#1#2{{$\null^{#1}$\if@tempswa\typeout
	{IJCGA warning: optional citation argument
	ignored: `#2'} \fi}}
\newcommand{\citeup}{\cite}

\def\fnm#1{$^{\mbox{\scriptsize #1}}$}
\def\fnt#1#2{\footnotetext{\kern-.3em
	{$^{\mbox{\sevenrm #1}}$}{#2}}}

\font\twelvebf=cmbx10 scaled\magstep 1
\font\twelverm=cmr10 scaled\magstep 1
\font\twelveit=cmti10 scaled\magstep 1
\font\elevenbfit=cmbxti10 scaled\magstephalf
\font\elevenbf=cmbx10 scaled\magstephalf
\font\elevenrm=cmr10 scaled\magstephalf
\font\elevenit=cmti10 scaled\magstephalf
\font\bfit=cmbxti10
\font\tenbf=cmbx10
\font\tenrm=cmr10
\font\tenit=cmti10
\font\ninebf=cmbx9
\font\ninerm=cmr9
\font\nineit=cmti9
\font\eightbf=cmbx8
\font\eightrm=cmr8
\font\eightit=cmti8


\centerline{\tenbf CHIRAL RANDOM MATRIX THEORY AND QCD}
\baselineskip=22pt
\vspace{0.8cm}
\centerline{\tenrm Jacobus Verbaarschot}
\baselineskip=13pt
\centerline{\tenit Department of Physics, SUNY at Stony Brook}
\baselineskip=12pt
\centerline{\tenit Stony Brook, NY 11790, USA}
\vspace{0.3cm}
\vspace{0.9cm}
\abstracts{As was shown by Leutwyler and Smilga, the fact that chiral
symmetry is broken and the existence of a effective finite
volume partition function leads to an infinite number of
sum rules for the eigenvalues of the Dirac
operator in QCD. In this paper we argue these constraints, together
with universality arguments from
quantum chaos and universal conductance fluctuations, completely
determine its spectrum near zero virtuality. As in the classical random
matrix ensembles, we find three universality classes, depedending on
whether the color representation of the gauge group is pseudo-real, complex
or real. They correspond to $SU(2)$ with fundamental fermions, $SU(N_c)$,
$N_c \ge 3$, with fundamental fermions, and $SU(N_c)$, $N_c \ge 3$, with
adjoint fermions, respectively.}

\vfil

\newcommand{\be}{\begin{eqnarray}}
\newcommand{\ee}{\end{eqnarray}}
\newcommand\del{\partial}
\newcommand\barr{|}
\newcommand\cita{\cite}
\newcommand\half{\frac 12}
\rm\baselineskip=14pt
\section{Introduction}

It is widely believed that QCD is $the$ theory of strong interactions
which describes the hadronic world as we know it. On the other hand, at low
energies, the same physics can be described
much more economically in terms of a partition function involving
effective degrees of freedom. This raises the question to which extent
the low energy effective theory put constraints on the underlying
microscopic theory.

This question can be answered if we restrict ourselves to a space time
volume $V_4 = L^4$ with $L \ll m_\pi^{-1}$. In that
case only the static modes
of the Goldstone fields contribute to the partition function. If, at the
same time, $L \gg \Lambda^{-1}_{QCD}$, the low-energy effective
paritition function coincides with the
QCD partition function\cite{LS}. Its mass dependence is given
by\cite{LS}
\be
Z^{eff}(m,\theta) = \int_{U\in G/H} DU \exp(V_4\Sigma \,{\rm Re} {\rm Tr}
\,me^{i\theta/N_f} U )
\label{1.1}
\ee
where the integral is over the coset $G/H$. For the standard scheme
of chiral symmetry breaking we have $G/H = SU(N_f)\times SU(N_f)/SU(N_f)$.
In agreement\cite{SHIFMAN-pr} with the Vafa-Witten
\cite{VAFA-WITTEN} theorem,
this corresponds to a condensate $\Sigma$ that is flavor symmetric
(see also refs.\cite{PESKIN,SMILGA-V}). For
simplicity we take the mass matrix $m$ diagonal in this paper. As dictated
by QCD Ward identities,
the masses  always occur in the combination $me^{i\theta/N_f}$, where
$\theta$ is the vacuum angle.

The mass dependence of the QCD partition function is given by
\be
Z^{QCD}(m,\theta) = \sum_\nu e^{i\nu\theta}
\langle \prod_f m_f^\nu\prod_{\lambda_n>0}(\lambda_n^2 + m^2_f) \rangle_\nu,
\label{1.2}
\ee
where the average $\langle \cdots \rangle_\nu$
is over all gauge configurations with topological charge
$\nu$ weighted by the  gluonic action.

As was argued correctly by Leutwyler and Smilga\cite{LS}, the
the finite volume partition function, imposes constraints on the
eigenvalues of the Dirac operator. More precisely, sum rules are obtained
by equating the coefficients of the expansion in powers of the quark masses
of both sides of the equation
\be
\frac{Z^{\rm QCD}_\nu(m)}{Z^{\rm QCD}_\nu(0)}
= \frac{Z^{\rm eff}_\nu(m)}{Z^{\rm eff}_\nu(0)},
\label{1.3}
\ee
where $Z_\nu(m)$ is the $\nu$'th Fourier coefficient of $Z(m,\theta)$.
Since this leads to an infinite set of sum rules, the following
question should be raised: to what extent is the spectrum of
the Dirac operator determined
by the finite volume partition function? Because we study a region
where only the $static$ modes contribute to the
finite volume partition function, we
expect that it only determines  the part of the spectrum $near$ $zero$
$virtuality$.
To make this statement more precise let us consider the simplest sum rule
obtained by equating the ${\cal O}(m^2) $ terms in (\ref{1.3})
\be
\frac 1{V_4^2}\left \langle {\sum}' \frac 1{\lambda_n^2} \right\rangle_\nu
 = \frac {\Sigma^2}{4(N_f + \nu)},
\label{1.4}
\ee
where the sum is over the positive nonzero eigenvalues of the Dirac operator.
If the spectral density is defined by
\be
\rho(\lambda) = \langle{\sum}' \delta(\lambda -\lambda_n)\rangle
\label{1.5}
\ee
we can rewrite the l.h.s of (\ref{1.4}) as
\be
 \frac 1{V_4^2}\langle {\sum}' \frac 1{\lambda_n^2} \rangle
= \int dz \frac 1{V_4} \rho(\frac{z}{V_4}) \frac 1{z^2},
\label{1.6}
\ee
where we have introduced a new integration variable by
\be
z = \lambda V_4.
\label{1.7}
\ee
This leads to the definition of the microscopic limit of the spectral
density\cite{SV-1993}
\be
\rho_S(z) = \lim_{V_4 \rightarrow \infty} \frac 1{V_4} \rho(\frac{z}{V_4}).
\label{1.8}
\ee
It is clear that information of the static finite volume partition function
is contained in this limiting function of the spectral density.
Since
$Z^{\rm eff}$ is based on symmetries and chiral symmetry breaking only we
arrive at the conjecture that the microscopic spectral density
is completely determined by the symmetries of the QCD partition function.
The same applies to microscopic correlation functions of the spectral
density which appear in higher order sum rules.

It is our program to contruct a spectral density that contains nothing else
but the symmetries of the QCD partition function as input.
Such question has been
phrased in general terms by Balian\cite{BALIAN-1968},
who introduced the maximal entropy
approach to obtain the probability distribution of the ensemble
that decribes a system that obeys a set of well defined constraints.
This approach has been applied successfully the the theory of statistical
$S$-matrix fluctuations\cite{MELLO,VWZ,VERB-ann}
and universal conductance fluctuations\cite{MELLO-univ,WEID-ZUK,BEENHAKKER}.
For example, in the case of the classical random matrix ensembles, a Gaussian
probabilty distribution of the matrix elements is obtained in this way.

Along these lines we can construct a random matrix model with the symmetries
of the QCD partition function as input. As in the classical random
matrix ensembles\cite{DYSON-three},
we find three different universality classes\cite{JJV-three}: 1. The chiral
orthogonal ensemble (chGOE) for $SU(2)$ with fundamental fermions when
the Dirac operator is real. The chiral unitary ensemble (chGUE) for
$SU(N_c)$, $N_c \ge 3$, with fundamental fermions (in this case the Dirac
operator is complex). 3. The chiral symplectic ensemble (chGSE) for
$SU(N_c)$, $N_c \ge 2$, with adjoint fermions (in this case the Dirac operator
can be regrouped into real quaternions). In all three models the scale
is set by the chiral condensate, which, via the Banks-Casher
formula\cite{BANKS-CASHER},
is related to the average level density. Further they
depend on the number of flavors and the total toplogical charge.
Precursers of the chGUE discussed in this paper can be found in
refs.\cite{NOWAK,SIMINOV}.

Of course, it makes only sense to use random matrix theory for observables
that are not sensitive to the detailed dynamics of the system. From
studies of nuclear level resonances\cite{PORTER,BOHIGAS},
universal conductance fluctuations\cite{STONE,WEIDI},
quantum chaos
\cite{BERRY,BOHIGAS-billiard,BOHIGAS-GIANNONI,SELIGMAN-VZ,SELIGMAN-V}
and general arguments\cite{Brezin}
we know that correlations between eigenvalues on the
scale of a finite number level spacings are such observables.
One of the best known examples
is the nearest neighbor spacing distribution. Because
the micrsocopic spectral density (\ref{1.8}) describes
correlations on the scale of a finite number of levels away from zero,
we expect that this quantity is universal.
It is our conjecture that the exact QCD microscopic spectral density
belongs to this universality class.

What are the arguments in favor of universality? In the ideal case
one would like to analyze the spectrum of the QCD-Dirac operator as obtained
in numerical lattice calculations. However, technically
it is not possible to work close enough to the chiral limit, and the best
one can do in this direction is to work with cooled configurations which
are represented by a partition function of a liquid of instantons and
anti-instantons\cite{chfluid} (see however\cite{CHRIST}).
In this paper we summarize the main
numerical results (see section 6).
Important support in favor universality comes
from a different branch of physics, namely from
the theory of universal conductance fluctuations. In that context, the
microscopic spectral density of the eigenvalues of the transmission matrix,
was calculated for the Hofstadter\cite{HOFSTADTER}
model, and, to a high degree of accuracy, it agrees with
the random matrix prediction\cite{SLEVIN-NAGAO}.
The Hofstadter model, which has the symmetries of the chGUE for zero
flavors, is therefore in the same universality class as the corresponding
random matrix model.  Because, this model differs in all
other respects from the random matrix model, this result illustrates
the size of the basin of attraction of the random matrix model and
greatly increases the hope that the exact QCD microscopic spectral
density also belongs to this universality class.

Other arguments in favor of the unversality come from the finite
volume partition function. Using general arguments used before in
ref.\cite{LS} for $SU(N_c)$, $N_c \ge 3$, with fundamental
fermions, it is possible
to write down the effective partition function for $SU(2)$ gauge theories
with fundamental fermions and for $SU(N_c)$ theories with adjoint fermions.
This allows us\cite{LS,SMILGA-V} to derive sum
rules for the inverse eigenvalues of the Dirac operator for an arbitrary
number of flavors, and for a given value of the topological charge.
All results obtained this way coincide with sum-rules derived
from random matrix theory\cite{V-Selberg}.

Questions related to the unversality of correlations in quantum spectra
are discussed in section 2. The symmetries of the QCD partition
function will be analyzed in section 3.
The main part of our program is
to construct a random matrix model that has nothing else but these
symmetries as input. This so called chiral random matrix model
will be introduced in section 4.
A summary of the analytical results will be
presented in section 5, and numerical results for the microscopic spectral
density in a liquid of instantons are shown in section 6. Concluding
remarks are made in section 7.

\section{Universality of Eigenvalue Correlations in Quantum spectra}

More than three decades ago it was realized that correlations between
the energy levels of compound nuclei show a universal behavior
(see for example ref.\cite{PORTER}). First, it
was discovered that the nearest neighbor spacing distribution follows the
so called Wigner surmise. Later it was found that not only short
range correlations but also correlations over many level spacing do not
depend on the specific dynamics of the system. One of the most outstanding
characteristics is the stiffness or rigidity of the spectrum. For instance,
if an energy interval contains on an average $N$ levels, the variance of
the distribution, $\Sigma_2(N)$,
is not $N$ (as for a random sequence), but rather
$(\log N)/\pi^2$. Both results can be derived from a random matrix theory
with only the symmetries of the system as input. This provides a strong
argument in favor of the universality of such level correlations.
In particular through the work of Dyson\cite{DYSON-three},
we know that there are three distinct universality classes corresponding to
real symmetric, complex hermitean or quaternion real hamiltonian matrices.
They correspond to systems with time reversal invariance and integral spin
or rotational invariance, with broken
time reversal invariance, and half-integer
spin systems with time reversal invariance and no rotational invariance.

Much more recently, it was realized that random matrix correlations
do not only appear in complex systems with many degrees of freedom, but
also in systems with as few as two degrees of freedom. This was first shown
by pioneering studies with quantum billiards\cite{BERRY,BOHIGAS-billiard}.
Later it was found that complete chaos is not only a sufficient ingredient
but also a necessary ingredient\cite{SELIGMAN-VZ}. As an illustration,
\cite{SELIGMAN-V}
we show in Fig.~1  the $\Delta_3(L)$ statistic
(an integral transform of the number variance
$\Sigma_2(N)$) versus the length of an interval containing on an average $L$
levels, and the nearest neighbor spacing spacing distribution,
$P(S)$, calculated from the (approximately) hundred lowest levels of
the Hamiltonian
\be
H = \frac 12((p_x - \frac B2 y)^2+ (p_y - \frac B2 x)^2) +
     \frac 12 ((x/\sigma_1)^6 + (y/\sigma_2)^6 + g(x-y)^6).
\label{1}
\ee
For a suitable choice of the parameters $\sigma_1$, $\sigma_2$ and the
magnetic field, this Hamiltonian is completely chaotic for $g= 0$.
Although in this case the time reversal symmetry is broken an additional
symmetry allows us to choose a basis in which the Hamiltonian is real. As shown
in Fig. 1 the spectral correlations are indeed
described by the GOE. Only when the coupling constant is large enough to
destroy all remnants of this symmetry,  while $B\ne 0$ is
kept at its original value (broken time reversal invariance),
the spectral correlations are given by the GUE (see Fig. 2).
\vskip 5cm
\fcaption{Numerical results for the $\Delta_3$ statistic (squares) and
the nearest neighbor spacing distribution (histogram) for a nonzero
magnetic field and $g = 0$.
The full and the dashed curves show the corresponding results for the
GUE and the GOE, respectively.}
\vskip 5cm
\fcaption{Numerical results for a nonzero magnetic field and $g\ne 0$.
See the caption of Fig. 1 for further explanation.}

Two more cases of universal correlations should be mentioned. First,
cross-section fluctuations in compound nuclei are determined by the average
$S$ matrix\cite{VWZ}.
Second, universal conductance fluctuations: the variance of the conductivity
is a given by a pure number times the quantum conductance $2e^2/\hbar$.
This phenomenon is closely related to  $S-$matrix fluctuations\cite{WEIDI},
in particular to Ericson fluctuations\cite{ERICSON}
which involve correlations between two $S$ matrices
at different energies.

What did we learn from the above studies?
One very important implicit ingredient in
all cases is the separation of scales:
a slow varying or average scale and a rapid varying or microscopic scale.
This allows us to express all correlations in terms
of the slow varying scale, the average level spacing, the average $S$-matrix,
or the average conductance. This would not be
not possible if fluctuations on all scales
were present. The central point is that only correlations expressed
in terms of the rescaled variables are universal and can be described by
random matrix theory.
In particular, for level correlations it means that only
correlations of the unfolded spectrum $\{\lambda_n^U\}$ defined by
($\bar{\rho}$ is the average level density)
\be
\lambda_n^U = \int_{-\infty}^{\lambda_n} \overline{\rho}(\lambda) d\lambda,
\label{2}
\ee
are described by random matrix theory.

Second, the microscopic correlations are completely determined
by the symmetries of the system.
In the QCD partition function, the eigenvalues show fluctuations
over the ensemble of gauge field configurations. From what is said above
we expect that only the microscopic correlations are universal and
can be described by random matrix theory. In the next section we will
analyze the symmetries of the QCD-Dirac operator and use that as an
input for the random matrix model to be constructed later.

\section{Symmetries of the Dirac operator}

In this section we consider the symmetries of the QCD Dirac operator
for a fixed  external gauge field $A_\mu$. The masses of the quarks are taken
to be zero (chiral limit). In that case
the Dirac operator $D$ is chirally symmetric
\be
\{ \gamma_5, D\} = 0.
\label{3}
\ee
As a consequence the eigenvalues occurs in pairs $\pm \lambda_n$, and
the eigenfunctions have opposite chirality. The only
exception are the zero eigenvalues. For field configurations with topological
charge $\nu$ we have exactly $\nu$ zero modes, all of the same chirality.

A less obvious symmetry, which plays an important role in random
matrix theory, is the symmetry that dictates the Wigner type of color
representation of the gauge group.
As suggested by the classical random matrix ensembles, the reality type of the
matrix elements determines the corresponding universality class.
In QCD, with three colors and
fundamental fermions, there is no additional symmetry and
 the color representation of the gauge group
is complex. Its Lagrangian
is invariant under $U(N_f) \times U(N_f)$ (a subgroup $U_A(1)$ is broken
by the quantum fluctuations). As is well known this symmetry group
is enlarged to $U(2N_f)$\cite{PESKIN,Shifman-three}
for $SU(2)$ with
fundamental fermions, and to $SU(N_f)$ for $SU(N_c)$ in the adjoint
representation with $N_f$ Majorana fermions.
This fact is closely related to the Wigner type of the
representation of the color group.

Indeed, for $SU(2)$ with fundamental fermions the Dirac operator,
\be
D = i \gamma_\mu \del_\mu + \gamma_\mu A_\mu^a \frac {\tau^a}2
\label{3.1}
\ee
has an
additional symmetry
\be
[i\gamma_2\gamma_4 \tau_2^{\rm color} K, D] = 0.
\label{3.2}
\ee
Here, $\tau_k$ are the Pauli matrices acting in color space, and $K$ is the
complex conjugation operator. The symmetry operator satisfies the property
\be
(i\gamma_2\gamma_4 \tau_2^{\rm color} K)^2 = 1,
\label{3.3}
\ee
which allows us to choose a basis in which the Dirac operator is real.
Note, however, that the representation of the color group is pseudo-real.
Because of the symmetry (\ref{3.2}) the symmetry group is enlarged to
$U(2N_f)$,

For gauge theories with adjoint fermions the Dirac operator is given
by
\be
D= i \gamma_\mu \del_\mu + i f_{klm} \gamma_\mu A_\mu^m,
\label{3.4}
\ee
where the $f_{klm}$ are the structure constants of the gauge group. In this
case the additional symmetry follows from the commutator
\be
[i\gamma_2\gamma_4 K, D] = 0.
\label{3.5}
\ee
Now, the symmetry operator satisfies
\be
(i\gamma_2\gamma_4 K)^2 = -1,
\label{3.6}
\ee
which allows us to choose a basis in which the Dirac matrix can be regrouped
such that its matrix elements are quaternion real. In this case the
representation of the color group is real. Because adjoint fermions are
formulated in terms of Majorana fermions, the symmetry (\ref{3.5}) results
in an $SU(N_f)$ flavor symmetry.

In each of the three cases discussed above the scheme of chiral symmetry
breaking is different. For $SU(N_c)$, $N_c \ge 3$, with fundamental
fermions the chiral symmetry is broken according to
\be
SU(N_f) \times SU(N_f) \supset SU(N_f).
\ee
For $SU(2)$ with fundamental fermions the symmetry group is enlarged
to $SU(2N_f)$ and spontaneously broken according to
\be
SU(2N_f) \supset Sp(2N_f).
\ee
In the case of adjoint fermions and an $SU(N_c)$, $N_c \ge 2$, color
group the flavor symmetry group is $SU(2N_f)$. In this case
the symmetry breaking pattern is
\be
SU(N_f) \supset O(N_f).
\ee

The different symmetry breaking schemes are determined by the dynamics of
the theory\cite{PESKIN}. They have to obey general constraints due
to the Vafa-Witten theorem, and, in agreement with general
QCD inequalities\cite{ineq}, they should give rise to pseudoscalar
Goldstone bosons. More discussion of this point can be found in
ref.\cite{SMILGA-V}.

\section{Chiral Random Matrix Theory}

The structure of the Dirac operator in QCD is much richer than that of a
Hamiltonian of a completely chaotic non-relativistic quantum system.
Nevertheless, it is possible to write down a random matrix theory that
includes all symmetries discussed in previous section.
For $N_f$ flavors the partition function for the sector of topological charge
$\nu$ is defined by
\be
Z_\nu = \int {\cal D}T \prod_{f=1}^{N_f}\det \left (
\begin{array}{cc} m_f & iT\\
                 iT^\dagger & m_f
\end{array} \right )
 \exp(-\frac{n\beta}{2\sigma^2} {\rm Tr }T^\dagger  T)
\label{4}
\ee
Here, the integral is over $n\times m$ matrices $T$ with
the Haar measure. The determinant in this equation plays the role of the
fermion determinant in QCD. We want to remark that the matrix
\be
\left ( \begin{array}{cc} 0 & iT\\
                 iT^\dagger & 0
\end{array} \right ).
\label{5}
\ee
has exactly $\nu = |n-m|$ zero eigenvalues. All other eigenvalues occur
in pairs $\pm \lambda_n$. We will identify $N=n+m$ as the volume of space time.
For definiteness we take $m> n$, and, as in QCD, we always take $\nu \ll N$.
The connection with the real world goes through the parameter $\Sigma$.
In this model it is given by
\be
\Sigma = \lim_{m\rightarrow 0} \lim_{N\rightarrow\infty} \frac{\pi \rho(0)}{N}.
\label{6}
\ee
According to the Banks-Casher formula\cite{BANKS-CASHER} $\Sigma$ can
be identified as the chiral condensate.

The representation of the color group is implemented via the the matrix
elements $T$ and the integration measure. The number of independent
variables per matrix element is denoted by $\beta$
(called Dyson parameter)\footnote{This parameter
has also been introduced in the exponent, so that
the average level density becomes independent of $\beta$.}.
For $SU(2)$ with fundamental
fermions the color representation is pseudoreal and the Dirac operator
is real. The integral in Eq. (\ref{4}) is over real matrices and we have
$\beta = 1$. For $SU(N_c)$, $N_c \ge 3$, both the color representation and
the Dirac operator are complex. The matrix $T$ in Eq. (\ref{4}) is complex
and $\beta = 2$. Finally, for $SU(N_c)$, $N_c \ge 3$, with adjoint fermions,
the color representation is real, and the Dirac operator is quaternion real.
The matrix elements of $T$ are also quaternion real, and we have $\beta = 4$.

An explicit realization of this partition function is for a liquid of
instantons. In that case the Dirac operator is evaluated in the space of
fermionic zero modes leading to the same matrix structure as in Eq.
(\ref{4}). However, instead of the integral over the matrix elements,
we have an integral over the collective coordinates of the instantons.
For a discussion and a numerical evaluation of this partition
function we refer to setion 7.

\section{Analytical Results}

The theorical analysis of the partition function Eq. (\ref{4}) is
straightforward. The eigenvalue distribution is obtained by changing
the integration variables according to
\be
T = U \Lambda V^{-1},
\label{5.1}
\ee
where $\Lambda$ is a diagonal matrix with positive real matrix elements.
The Jacobian of this transformation  is\cite{JJV-three}
\be
J(\Lambda) = \prod_{k<l} |\lambda_k^2 -\lambda_l^2|^\beta \prod_k
\lambda_k^{\beta\nu +\beta -1}.
\label{5.2}
\ee
This Jacobian can be obtained by an explicit calculation
(see\cite{HUA} for a discussion
of the necessary techniques). It also follows
from the fact that the Jacobian vanishes when two eigenvalues
coincide or one eigenvalue equals zero, and that it should depend
symmetrically on all eigenvalues. The correct powers then follow from
dimensinal arguments.

The integration over $U$ and $V$ only contributes a constant factor, so that
the parition function is given by
\be
Z_\nu = C_{\beta, n}\int d \lambda_1\cdots d\lambda_n
\prod_{k<l} |\lambda_k^2 -\lambda_l^2|^\beta \prod_{k}
\lambda_k^{2N_f +\beta\nu+\beta -1}
\exp\left(-\frac{n\beta\Sigma^2}{2} \sum_k \lambda^2_k\right),
\label{5.3}
\ee
where the constant $C_{\beta, n}$ is determined by the normalization.
The joint eigenvalue distribution $\rho(\lambda_1,\cdots, \lambda_n)$ is
just the integrand of Eq. (\ref{5.3}). We have
\be
Z = \int d \lambda_1\cdots d\lambda_n \rho(\lambda_1,\cdots, \lambda_n).
\label{5.4}
\ee
The eigenvalue distribution follows by integrating over all eigenvalues
except one
\be
\rho(\lambda_1) = \int d \lambda_2\cdots d\lambda_n \rho(\lambda_1,\cdots,
\lambda_n).
\label{5.5}
\ee

Integrals of this type have been widely studied in the context of random
matrix theory. One very powerful method is the orthogonal polynomial method
Dyson, Wigner and Mehta (see the book of Mehta\cite{MEHTA} for references).
This method is particularly suited for $\beta = 2$,
when the integrals follow immediatly from the orthogonality relations
of the orthogonal polynomials. For $\beta = 1$ and $\beta = 4$ it is still
possible to do the integrals, but the resulting expressions are much more
complicated. Instead of classical orthogonal polynomials, one has to
invoke skew orthogonal polynomials. The general method was developed
by Dyson\cite{DYSON-skew} and by Mahoux and Mehta\cite{MAHOUX-MEHTA},
and applied to many different ensembles
by Nagao, Slevin and Wadati\cite{NAGAO-SLEVIN}.
Unfortunately, the generalized Laguerre ensemble for $\beta =1$
was not analyzed before, so we perfomed the analysis in ref.\cite{V_GOE}.
The symplectic $generalized$ Laguerre ensemble has not yet been studied, but
using similar methods it is possible to obtain an explicit expression for
the level density.

First we discuss the simplest case, $\beta = 2$. This case is also known
as the unitary generalized Laguerre ensemble and was studied
in\cite{KAHN,BRONK,TRACY,FORRESTER,NAGAO-SLEVIN}.
The latter authors also evaluated the microscopic limit.
This ensemble was put in the context of QCD in
ref.\cite{VERBAARSCHOT-ZAHED-1993},
where the microscopic spectral density for $\nu = 0 $ was derived. The
microscopic spectral density for arbitrary $\nu$ was first given
in ref.\cite{V_GOE}.

The result for the microscopic spectral density (\ref{1.8})
for the chGUE ($\beta = 2$) is\cite{VERBAARSCHOT-ZAHED-1993,V_GOE}
\be
\rho_S(z)  = \frac {\Sigma^2 z}{2} (J^2_{N_f+\nu}(\Sigma z) -
J_{N_f+\nu+1}(\Sigma z) J_{N_f+\nu-1}(\Sigma z)).
\label{5.6}
\ee
It depends only on the combination $N_f +\nu$ which agrees with the sum rules
derived by Leutwyler and Smilga\cite{LS}. It is also possible
to obtain explicit expressions for the two level correlation functions.
For results we refer to the literature\cite{VERBAARSCHOT-ZAHED-1993}.

The expressions for  $SU(2)$ with fundamental fermions ($\beta =1$) are much
more complicated. In this case we find
the microscopic spectral density\cite{V_GOE}
\be
\rho_S(z) = \frac {\Sigma}{4} J_{2a+1}(z{\Sigma}) &+& \frac {\Sigma}{2}
\int_0^\infty dw (zw)^{2a+1} \epsilon(z-w)
\left ( \frac 1w \frac d{dw} - \frac 1z \frac d{dz}\right )\nonumber \\
&\times&
\frac{wJ_{2a}(z{\Sigma})J_{2a-1}( w{\Sigma})
-zJ_{2a-1}(z{\Sigma})J_{2a}(w{\Sigma})}{(zw)^{2a}(z^2-w^2)},
\label{5.7}
\ee
where $a$ is the combination
\be
a = N_f -\frac 12 +\frac {\nu}2.
\label{5.8}
\ee

Eqs. (\ref{5.6}) and (\ref{5.7}) can be used to derive Leutwyler-Smilga
sum rules. However, there is a more direct way to obtain these results.
Exactly, these type of integrals
follow from the Selberg formula\cite{SELBERG,MEHTA}.
Sum rules have been obtained\cite{V-Selberg} for
the quantities
\be
S_p \equiv \left <\sum_{n_1\ne n_2\ne\cdots\ne n_p}
\frac 1{\lambda_{n_1}^2\cdots \lambda_{n_p}^2} \right >,
\label{5.9}
\ee
and
\be
S_{pq} \equiv \left < \sum_{{\rm all}\,\, \lambda_i\,\, {\rm different}}
\frac 1{\lambda_{m_1}^4
\cdots \lambda_{m_p}^4\lambda_{n_1}^2\cdots\lambda_{n
_q}^2}\right>.
\label{5.10}
\ee
The integrals can be calculated by deriving a recursion relation.
It is also possible to derive recursion relations for more complicated
sum rules.
However, then the recursion branches off and cannot be solved
analytically any longer.

The results for the simplest family  of sum rules are given by
\be
S_p = \frac 1{\prod_{k=0}^{p-1} ({\beta +\beta\nu}+ 2(N_f-1) +\beta k)}
\left ( \begin{array}{c} n \\ p \end{array} \right )
\left( {\beta n\Sigma^2} \right )^p,
\label{5.11}
\ee
which, in the limit $N\rightarrow\infty$, simplifies to
\be
S_p = \frac{(N^2\Sigma^2)^p}{2^{2p} p!} \frac{\Gamma(\nu+1+2(N_f-1)/\beta)}
{\Gamma(\nu+p+1+2(N_f-1)/\beta)}.
\label{5.12}
\ee
This result for $\beta= 2$, and for $\beta=4$,  $p = 1$ and
$N_f =1 $ or $N_f = 2$, were derived previously with the help of the finite
volume static partition function\cite{LS}.
Results for all other case, in particular
for $SU(2)$ with fundamental fermions, were first derived with the help of
random matrix theory\cite{V-Selberg}. Presently, we have shown
that the $p =1$ sum rules can also be obtained starting from the
finite volume partition function\cite{SMILGA-V}.

Note, that for $N_f =1 $ the parameter
$\beta$ drops from the equation. Indeed, in this
case the finite volume partition function is solely determined by the
$U_A(1)$ anomaly and is not dependent on the Wigner type of the color
representation. There are no additional degrees of freedom that, for
more than one flavor, contain information on the color representation.
This implies that the microscopic limit of the spectral density
is not determined uniquely by the finite volume partition function.

The results for $S_{pq}$ can be derived in a similar fashion. We only
quote the result\cite{V-Selberg}
\be
S_{pq} = N_{pq}\frac {  \Gamma(\alpha + \frac 2{\beta}) \Gamma(\alpha+p)
                        \Gamma(\alpha + \frac 2{\beta} + p + n)}
                     {\Gamma(\alpha + \frac 2{\beta}+q+2p) \Gamma(\alpha)
                        \Gamma(\alpha + \frac 2{\beta} + n)}
                   \left (  n^2\Sigma^2 \right )^{q+2p},
\label{5.13}
\ee
where the factor $N_{pq}$ is defined by
\be
N_{pq}=\frac {n!}{p! q! (n-p-q)!}
\ee
and
\be
\alpha = \nu + 1 + \frac{2(N_f -2)}{\beta}.
\label{5.14}
\ee
These sum rules depend on $\beta$. Indeed, the effective low energy field
theories are different for different values of $\beta$.
Even for $N_f =1$ they depend on $\beta$ (the
integrals should be convergent).
Since such sum rules cannot be derived from the finite volume partition
function these results are completely consistent.

The result (\ref{5.13}) was derived from the finite volume partition
function\cite{LS} only for $\beta =2$, $p =1$ and $q= 0$.

\section{Numerical Results}

It would be very interesting to check the results of section 5 by
lattice QCD simulations. However, in practise,
it is not possible to obtain a sufficient number of eigenvalues of the
Dirac operator close to the chiral limit (see however ref.\cite{CHRIST}, where
the spectrum near zero virtuality has been obtained for quark masses that are
much larger than the smallest eigenvalues, and in the present context,
corresponds to $N_f = 0$).

Now, remember that the correlations discussed above are based a random
Dirac operator. When we consider the fluctuations of the eigenvalues of the
Dirac operator over the ensemble  of the gauge field configurations, it
seems clear that if the random matrix results are reproduced by cooled
configurations, they are certainly found for the original uncooled
configurations. The sum over all cooled configurations is well known: it
is a dilute gas or liquid of instantons.
Although, it does not have confinement, the bulk of
hadronic physics can be reproduced by this model\cite{DP}.

The partition function for a
liquid of instantons is given by\cite{CDG,DP}
\be
Z = \int \prod_{i=1}^N d \Omega_i
\prod_f^{N_f}\det(T^\dagger T + m_f^2)
\exp(-S_{\rm glue}),
\label{6.1}
\ee
where the integral is over the collective coordinates of each instanton,
and the Dirac opertor is calculated in the space of fermionic zero modes
with matrix elements denoted by $T_{ij}$.
The gluonic action, $S_{\rm glue}$, contains the interaction
between the instantons approximated by a sum over all pairs. The total
number of instantons in $N$. In general, the number of instantons
is not equal to the number of anti-instantons. However, in the numerical
calculations to be presented below, this will be the case.
Further details of this model and the calculations presented in this section
can be found in ref.\cite{chfluid}.

\vskip 11cm
\fcaption{A histogram of the total spectral density of the Dirac operator
in a liquid of instantons}

In the partition function (\ref{6.1}) it is
straightforward to change the number
of colors from $N_c = 2$ to $N_c = 3$.
In fact, we use the same computer program that
covers both cases. At present it is not yet clear how to write down an
instanton partition function for adjoint fermions. For a discussion of the
difficulties related to this issue we refer to ref.\cite{LS}.

The numerical results have been obtained for 32 instantons and 32
anti-instantons confined to a box of
$2.37^3 \times 4.74 \Lambda_{QCD}^{-4}$. For $N_f= 0$, the average was over
100,000 configurations. For all other flavors the average was over only
20,000 (correlated) configurations. A histogram of the results
for the total spectral density is shown in Fig. 3. The results are for
two and three colors and one, two and three flavors (see label of the figure).
In Fig. 4 the
region of the spectrum near zero virtuality is put under the microscope.
We show
a histogram of the unfolded spectrum in terms of the microscopic variable $z$,
so that the average spacing between the eigenvalues is one.

\vskip 11cm
\fcaption{The microscopic spectral density. The full line reprsents the
numerical results. Random matrix result of the chGOE and the chGUE is shown by
the dotted and the dashed curves, respectively.}

In the same
figures we also show the random matrix results for the chGOE (dotted curve)
and the chGUE (dashed curve). The only deviation is that the oscillations
in the chGUE random matrix results seem not to be completely reproduced.
At the moment it is not clear whether that this is a real effect, of that
it is a consequence of the finite size of the system.

\section{Conclusions}

We have argued is that the average miscroscopic spectral density
of the QCD Dirac  operator follows from its
symmetries and a maximum entropy principle. The latter principle
uniquely determines the random matrix theory obeying all constraints
imposed by the symmetries. At present, it is a conjecture that the
the fluctuations of the eigenvalues near zero virtuality over the
ensemble of gauge field configurations as weighted by the QCD partition
function belong the the same universality class as the corresponding
chiral random matrix theory. We have presented several strong
arguments in favor of this conjecture. First, results from the
Hofstadter model show that the random matrix model
has a wide basin of attraction.
Second, the triality of the classical random matrix ensemble also
occurs in the chiral random matrix ensembles and coincides with the
different schemes of chiral symmetry breaking. Third, numerical
calculations for instantons agree with the random matrix result.
Fourth, all spectral sum rules that
can be obtained from the finite volume static partition function,
which coincides with the full QCD partition function in its region
of applicability, are also
given by the chiral random matrix theory.

The finite volume static partition function alone does not
determine the microscopic spectral
density uniquely. The most striking example is that of one flavor.
In that case the effective theory does not depend on the color representation,
whereas the microscopic spectral density is different in each of the
three universality classes.

To address the question which effective field theory contains the information
of the full microscopic spectral density, remember that the resolvent is
defined by
\be
G(z) = \left\langle \sum_{\lambda_n} \frac 1{z-\lambda_n}\right\rangle.
\label{7.2}
\ee
It can be expressed as follows in terms of the partition function
\be
i G(iz) = \left . \frac{d}{dz}\right |_{u=z} \left \langle \frac{
\prod(\lambda_n^2 +z^2)}{\prod(\lambda_n^2 +u^2)}
\prod_{f =1}^{N_f}(\lambda_n^2 +m^2_f)\right \rangle.
\label{7.3}
\ee
In (\ref{7.2}) and (\ref{7.3}) the sum and product is over the positive
eigenvalues only. Therefore, the spectral density for $N_f$ flavors
can be obtained from a partition function of $N_f + 1$ fermionic
flavors and one complex scalar flavor.

\section{Acknowledgements}
The reported work was partially supported by the US DOE grant
DE-FG-88ER40388. We acknowledge the NERSC at Lawrence Livermore where
most of the computations presented in this paper were performed.
E. Shuryak, A. Smilga and I. Zahed are thanked for useful discussions.
The author also
thanks the organizers of this workshop for their hospitality and for the
stimulating atmosphere of this meeting.


\section{References}

\begin{thebibliography}{9}

\bibitem{LS}
H.~Leutwyler and A.~Smilga,
Phys. Rev. {\bf D46} (1992) 5607.

\bibitem{VAFA-WITTEN}
C. Vafa and E. Witten, Nucl. Phys. {\bf B 234} (1984) 173.

\bibitem{SHIFMAN-pr}
M. Shifman, private communication (see ref.$^5$).

\bibitem{PESKIN}
M. Peskin, Nucl. Phys. {\bf B175} (1980) 197.

\bibitem{SMILGA-V}
A. Smilga and J. Verbaarschot, {\it Spectral sum rules and finite volume
partition function in gauge theories with real and pseudoreal fermions},
preprint SUNY-NTG-94/18, TPI-MINN-94/10-T (1994).

\bibitem{SV-1993}
E.~Shuryak and J.~Verbaarschot,
 Nucl. Phys. {\bf A560} (1993) (306).

\bibitem{BALIAN-1968}
R.~Balian,
Nuov. Cim. {\bf 57} (1968) 183.

\bibitem{MELLO}
P. Mello, P. Pereyra and T. Seligman, Ann. Phys. (N.Y.) {\bf 161} (1985) 254.

\bibitem{VWZ}
J.~Verbaarschot, H.~Weidenm{\"u}ller, and M.~Zirnbauer,
Phys. Rep. {\bf 129} (1985) 367.

\bibitem{VERB-ann}
J. Verbaarschot, Ann. Phys. (N.Y.) {\bf 168} (1986) 368.

\bibitem{MELLO-univ}
P. Mello, Phys. Rev. Lett. {\bf 60} (1988) 1089;
P. Mello, P. Pereyra and N. Kumar, Ann. Phys. (N.Y.) {\bf 181} (1988) 290.

\bibitem{WEID-ZUK}
S. Iida, H. Weidenm\"uller and J. Zuk, Ann. Phys. (N.Y.) {\bf 200} (1990) 219.

\bibitem{BEENHAKKER}
C. Beenhakker, Phys. Rev. Lett. {\bf 70} (1993) 1155.

\bibitem{DYSON-three}
F.J. Dyson,
J. Math. Phys. {\bf 3} (1962) 1199.

\bibitem{JJV-three}
J. Verbaarschot, Phys. Rev. Lett. (1994) (in press).

\bibitem{BANKS-CASHER}
T.~Banks and A.~Casher,
Nucl. Phys. {\bf B169} (1980) 103.

\bibitem{NOWAK}
Phys. Lett. {\bf 217B}, 157 (1989); Phys. Lett. {\bf 226B}, 382 (1989).

\bibitem{SIMINOV}
Y.~A. Simonov, Phys. Rev. {\bf D43} (1991) 3534.

\bibitem{PORTER}
C.E.~Porter,
'{\it Statistical theories of spectra: fluctuations}', Academic
Press, 1965.

\bibitem{BOHIGAS}
R. Haq, A. Pandey and O. Bohigas, Phys. Rev. Lett. {\bf 48} (1982) 1086.

\bibitem{STONE}
A. Stone, P. Mello, K. Muttalib and J. Pichard in {\it Mesoscopic Phenomena
in Solids} B. Altshuler, P. Lee and R. Wedd (eds.) (North-Holland,
Amsterdam, 1991)

\bibitem{WEIDI}
H. Weidenm\"uller,
in {\it Proceedings of T. Ericson's 60th birthday}.


\bibitem{BERRY}
M. Berry, Ann. Phys. (N.Y.) {\bf 131} (1981) 91.

\bibitem{BOHIGAS-billiard}
O.~Bohigas, M.~Giannoni, and C.~Schmit,
Phys.Rev.Lett. {\bf 52}, 1 (1984).

\bibitem{BOHIGAS-GIANNONI}
O.~Bohigas, M.~Giannoni,
in '{\it Mathematical and computational methods in
nuclear physics}', J.S. Dehesa et al. (eds.), Lecture notes in Physics
{\bf 209}, Springer Verlag 1984, p. 1;
B.Simons, A. Szafer and B. Altschuler, {\it Universality in quantum chaotic
spectra}, MIT preprint (1993);

\bibitem{SELIGMAN-VZ}
T.~Seligman, J.~Verbaarschot, and M.~Zirnbauer,
Phys. Rev. Lett. {\bf 53}, 215 (1984).

\bibitem{SELIGMAN-V}
T.~Seligman and J.~Verbaarschot, Phys. Lett. {\bf 108A} (1985) 183.

\bibitem{Brezin}
E. Br\'ezin and A. Zee,
Nucl. Phys. {\bf B402} (1993) 613;

\bibitem{chfluid}
J. Verbaarschot, {\it Spectrum of the Dirac operator in a QCD instanton Liquid:
two versus three colors}, preprint SUNY-NTG-94/6 (1994).

\bibitem{CHRIST}
N. Christ and S. Chandrasekharan, this volume.

\bibitem{HOFSTADTER}
D.~Hofstadter,
Phys. Rev. {\bf B14} (1976) 2239.

\bibitem{SLEVIN-NAGAO}
K.~Slevin and T.~Nagao,
Phys. Rev. Lett. {\bf 70} (1993) 635.

\bibitem{V-Selberg}
J. Verbaarschot, Phys. Lett. {\bf B} (1994) (in press).

\bibitem{ERICSON}
T. Ericson,
Phys. Rev. Lett. {\bf 5} (1960) 430.

\bibitem{ineq}
D. Weingarten, Phys. Rev. Lett. {\bf 51} (1983) 1830;
E. Witten, Phys. Rev. Lett. {\bf 51} (1983) 2351;
S. Nussinov and M. Spiegelglas, Phys. Lett. {\bf B181} (1986) 134.

\bibitem{HUA}
L.~Hua,
{\it Harmonic Analysis of Functions of Several Complex Variables in
  the Classical Domain}, tr. L. Ebner and A. Kor\'ani, American Mathematical
  Society, Providence, RI, 1963.

\bibitem{Shifman-three}
S. Dimopoulos, Nucl. Phys. {\bf B168} (1980) 69;
M. Vysotskii, Y. Kogan and M. Shifman,
Sov. J. Nucl. Phys. {\bf 42} (1985) 318;
D.I. Diakonov and V.Yu. Petrov,
in '{\it Quark Cluster Dynamics}', Proceeding of the 99th WE-Heraeus
Seminar, Bad Honnef, 1992 (eds. K. Goeke $et$ $al.$), Springer 1993.

\bibitem{MEHTA}
M.~Mehta,
{\it Random Matrices}, Academic Press, San Diego, 1991.

\bibitem{DYSON-skew}
F. Dyson, J. Math. Phys. {\bf 13} (1972) 90.

\bibitem{MAHOUX-MEHTA}
G. Mahoux and M. Mehta, J. Phys. I France {\bf I} (1991) 1093.

\bibitem{NAGAO-SLEVIN}
T. Nagao and K. Slevin, J. Math. Phys. {\bf 34} (1993) 2317;
J. Math. Phys. {\bf 34} (1993) 2075;
T. Nagao and M. Wadati, J. Phys. Soc. Japan {\bf 60} (1991) 2998;
J. Phys. Soc. Japan {\bf 62} (1993) 3845.

\bibitem{V_GOE}
J. Verbaarschot, {\it The spectrum of the Dirac operator near zero virtulity
for $N_c = 2$ and chiral random matrix theory}, preprint SUNY-NTG-94/2.

\bibitem{KAHN}
D. Fox and P. Kahn, Phys. Rev. {\bf 134} (1964) B1151.

\bibitem{BRONK}
B. Bronk, J. Math. Phys. {\bf 6} (1965) 228.

\bibitem{TRACY}
C. Tracy and H. Widom,
{\it Level spacing distributions and the Bessel kernel}, University
of California preprint ITD 92/93-7 (1992).

\bibitem{FORRESTER}
P. Forrester, Nucl. Phys. {\bf B[FS]402} (1993) 709.

\bibitem{VERBAARSCHOT-ZAHED-1993}
J.~Verbaarschot and I.~Zahed, Phys. Rev. Lett. {\bf 70} (1993) 3852.

\bibitem{SELBERG}
A. Selberg,
Norsk Mat. Tid. {\bf 26} (1944) 71.


\bibitem{CDG}
C. Callan, R. Dashen and D. Gross, Phys. Rev. {\bf D17} (1978) 2717.

\bibitem{DP}
D.I.~Diakonov and V.Yu.~Petrov, Nucl. Phys. {\bf B245} (1984) 259;
E. Shuryak and J. Verbaarschot, Nucl. Phys. {\bf B341} (1990) 1;
Nucl. Phys. {\bf B 410} (1993) 55.

\end{thebibliography}
\end{document}